\documentclass[traditabstract]{aa} 
\usepackage{graphicx}
\usepackage{subfigure}
\usepackage{natbib}
\begin{document}
\title{Detecting the Collapse of Cooperation in Evolving Networks}
\author{Matteo Cavaliere
      \inst{1}
      \and
      Guoli Yang
      \inst{1}
      \and
      Vincent Danos
      \inst{1}
      \and
      Vasilis Dakos
      \inst{2}             
}

\institute{School of Informatics, University of Edinburgh, Scotland, United Kingdom
  	\and
    Institute of Integrative Biology, Center for Adaptation to a Changing Environment, ETHZ, Zürich, Switzerland,  \email{vasilis.dakos@env.ethz.ch}
}

\abstract{The sustainability of structured biological, social, economic and ecological communities are often determined by the outcome of social conflicts between cooperative and selfish individuals (cheaters). Cheaters avoid the cost of contributing to the community and can occasionally spread in the population leading to the complete collapse of cooperation. Although such a collapse often unfolds unexpectedly bearing the traits of a critical transition, it is unclear whether one can detect the rising risk of cheater's invasions and loss of cooperation in an evolving community. Here, we combine dynamical networks and evolutionary game theory to study the abrupt loss of cooperation as a critical transition. We estimate the risk of collapse of cooperation after the introduction of a single cheater under gradually changing conditions. We observe a systematic increase in the average time it takes for cheaters to be eliminated from the community as the risk of collapse increases. We detect this risk based on changes in community structure and composition. Nonetheless, reliable detection depends on the mechanism that governs how cheaters evolve in the community. Our results suggest possible avenues for detecting the loss of cooperation in evolving communities.}


\maketitle

%

\section{Introduction} \label{introduction}

The sustainability of many biological, social, economic, and ecological communities is determined by the interplay between individual actions and collective dynamics \citep{levin1999fragile}. The successful performance of a community is often based on the cooperative attitude of individuals that pay a personal cost to distribute general benefits \citep{nowak2006five}. Nonetheless, although cooperation favors in general the success of a community, it can also facilitate the appearance of {\em cheaters} who take advantage of {\em cooperators}, spread in the community, and may even cause its collapse \citep{levin1999fragile}. The failure of cooperation in the presence of cheaters has been observed in many systems at different scales. For instance, cooperating cells of {\em Pseudomonas fluorescens} build biofilms that help them to grow better while mutant cells (cheaters) - that do not produce similar adhesive factors - take advantage of the existing structure in order to spread and to eventually cause the colony to collapse \citep{rainey2003evolution,popat2012quorum}. Fruiting bodies formed under starvation by cooperative cells of {\em Myxococcus xanthus} can also be invaded by cheaters leading to the disruption of the fruiting body structure and forcing the cooperative survivors to reinvest in reconstruction \citep{travisano2004strategies}. At a different scale and in a different context shifts in cooperation and cheating have been debated to be causes of economic crises, \citep{haldane2009rethinking} as well as causes for the "tragedy" of common pool resources, (e.g., fisheries, forests, etc.) \citep{ostrom2005understanding, bowles2011cooperative}. In all these systems a long-standing question has been to understand the mechanisms that allow cooperators to resist the reproductive advantage of selfish cheating individuals \citep{nowak2006five}. Among the many theoretical and experimental studies on the maintenance of cooperation \citep{nowak2006five, ohtsuki2006simple, rand2011dynamic, sanchez2013feedback}, scenarios where strategies co-evolve with population structure \citep{perc2010coevolutionary} are of particular interest as they show how structural properties in the population can affect the evolution of cooperation \citep{perc2010coevolutionary, wardil2014origin, rand2011dynamic, sanchez2013feedback}. For instance, it has been shown that not only the number of cheaters in the community is important, but also how and to whom cheaters are connected \citep{cavaliere2012prosperity}. The interplay between the way individuals are connected and the overall prosperity of a system \citep{bascompte2010structure, levin1999fragile} endogenously determines either the formation or the sudden collapse of cooperative communities \citep{levin1999fragile}. Despite our relative good understanding of the conditions that promote the failure of cooperation in a community, it is still difficult to  predict whether the appearance of a cheater will cause the eventual loss  of cooperation \citep{ostrom2005understanding,bowles2011cooperative}. This is because it is hard to identify the underlying conditions that increase the risk of collapse in practice. Thus, it is crucial to develop alternative ways for detecting the risk of collapse of cooperation in a structurally evolving community. 

Recent work has suggested that generic patterns in the dynamics of a system can be used to infer proximity to abrupt and unexpected changes termed critical transitions \citep{scheffer2009early}. These dynamical patterns are generic in the sense that they do not depend on the particular system in question, but they are determined by the mathematical phenomenon of critical slowing down (CSD) that occurs prior to local bifurcation points \citep{strogatz2014nonlinear, wissel1984universal}. A bifurcation point represents a threshold where a qualitative change in the equilibrium of a system takes place: the iconic case is the shift between two alternative equilibria at a crossing of a fold bifurcation. Close to a local bifurcation, CSD implies that the system takes longer to recover back to equilibrium after a disturbance \citep{scheffer2009early, van2007slow}. In addition, the dynamics of the system become strongly variable \citep{carpenter2006rising}, and highly autocorrelated \citep{held2004detection}. As such, decreasing recovery rate, rising variance, and rising autocorrelation can all be used as early-warning signals of approaching critical transitions \citep{scheffer2009early, dakos2012methods}, or more general as indicators of loss of resilience \citep{dakos2014critical}. CSD indicators have been identified in a variety of systems: from the collapse of cyanobacteria \citep{veraart2012recovery}, zooplankton \citep{drake2005population}, and yeast populations \citep{dai2012generic} in the lab, to changes in trophic structure in lakes \citep{carpenter2006rising}, as well as prior to abrupt past climatic events \citep{dakos2012methods}, and the onset of depression in human patients \citep{van2014critical}. Despite limitations and challenges in detecting resilience indicators \citep{dakos2015resilience, boettiger2012quantifying}, growing evidence supports their potential use across different disciplines\citep{scheffer2012anticipating}. The application of such indicators for detecting abrupt transitions in structurally complex communities is still, however, largely unexplored. There are few studies that have highlighted the emergence of tipping points in mutualistic communities of plants and their pollinators \citep{lever2014sudden}, and the detection of transitions in ecological\citep{dakos2014critical} and socio-ecological networks \citep{suweis2014early}. Nonetheless, these studies assume a static structure that does not allow changes in the interactions among network components. 

Here, for the first time to our knowledge, we combine evolutionary game theory and dynamical networks to detect the collapse of cooperation in an evolving (structurally dynamic) community, using resilience indicators for critical transitions. To do this, we adapt a network model that displays cooperation collapses as consequence of cheater's invasions\citep{cavaliere2012prosperity}. Evolution of cheaters and cooperators is based on social inheritance: newcomers copy strategies and connections of successful role-models. Increased ability of newcomers to link to a high number of individuals already present in the network allows high prosperity but at the risk of facilitating cheater's invasions that can lead to the collapse of cooperation \citep{cavaliere2012prosperity}. We show that the collapse of cooperation occurs in an abrupt non-linear way that resembles a critical transition. Prior to collapse we estimate a series of indicators (structural and non-structural) that can signal the increasing fragility of cooperation. Our contribution is twofold. First, we demonstrate how the loss of cooperation in evolving communities bears the features of a critical transition. Second, we develop a set of structural-based indicators and state-based indicators for detecting approaching transitions in structurally evolving communities.

\section{Results}
\subsection{The Collapse of Cooperation in a Dynamical Network}\label{transitions}

\begin{figure}[!htb]
	\begin{center}
		\includegraphics[width=1\linewidth]{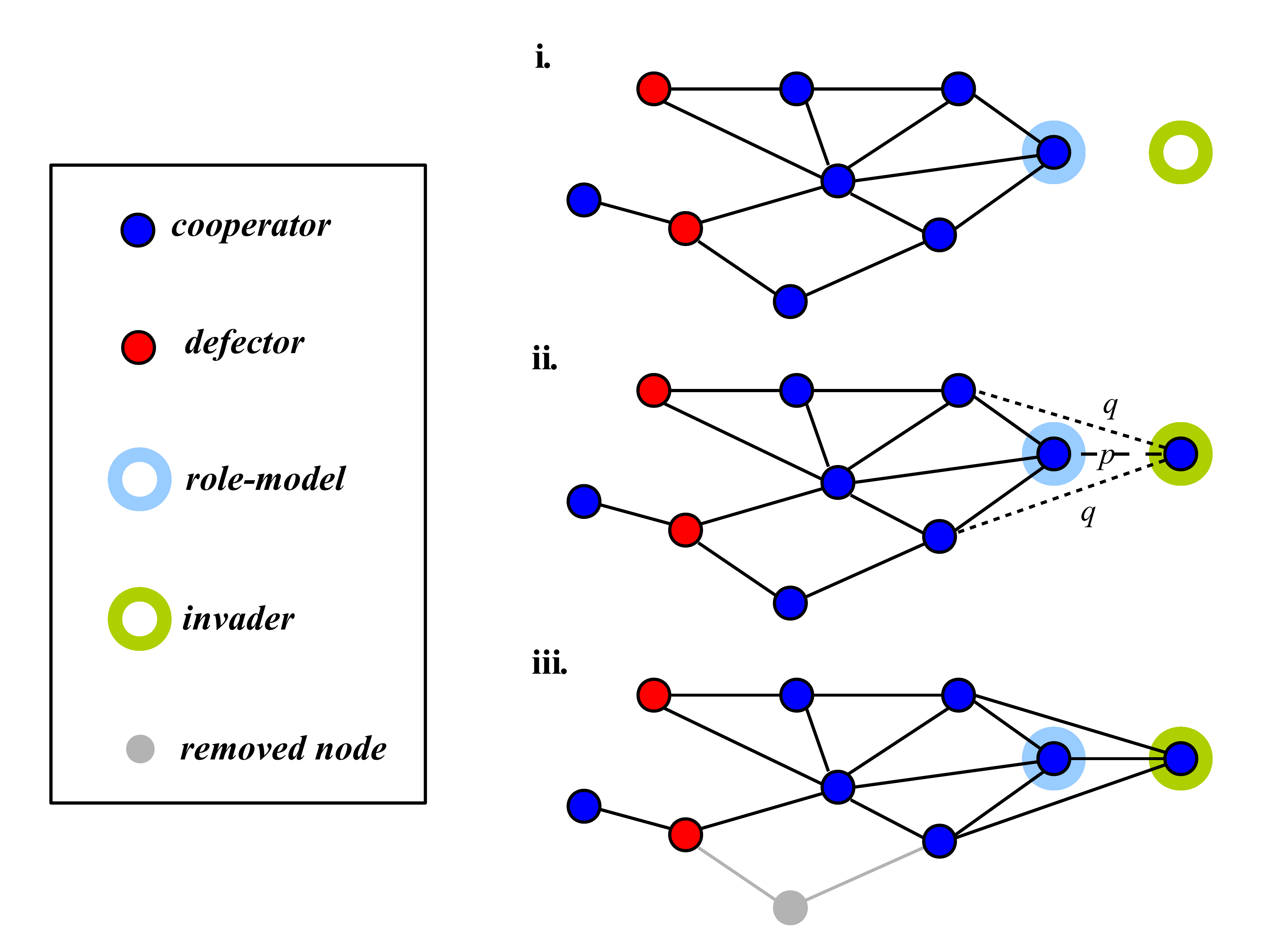} 
	\end{center}
	\caption{{\bf Evolution in a dynamical network.} The dynamics follow three distinct update steps in the considered model: 
		(step i) A role-model is selected based on its effective payoff.
		(step ii) The invader (newcomer) connects to the role-model with a probability $p$ (dashed line), connects to each of its neighbours with a probability $q$ (dotted lines) and emulates its strategy. $p$ and $q$ are called embedding parameters.
		(step iii) A randomly selected node and all its connections are removed from the network.}
	\label{fig:evolution}
\end{figure}

Following the work \citep{cavaliere2012prosperity} we consider a network model with a fixed number of agents (nodes) but with a non-fixed number of links: to whom and to how many neighbours an agent is connected varies during the evolution of the system. Each agent in the network can adopt one of the two strategies of the {\em Prisoner's Dilemma}. A {\em cooperator} pays a cost $c$ to provide a benefit $b$ to all of its neighbours. {\em Cheaters} pay no cost and distribute no benefit. For instance, if a cooperator has $m$ cooperative neighbours and $n$ cheating neighbours, its payoff is $m(b-c)-nc$. A cheater in the same neighbourhood has payoff $mb$. The dynamics are defined by a discrete sequence of {\em update steps} (Figure \ref{fig:evolution}). At each update step a newcomer ({\em invader}) is added and a randomly chosen existing node is removed so that the number of nodes is constant as $N$. As the newcomer has no specific strategy, a node $i$ is selected as a role-model with a probability proportional to its \emph{effective payoff} $EP_i$ = (1+$\delta)^{\emph{Payoff}_i}$, where $\delta \geq 0$ specifies the strength of selection and $\emph{Payoff}_i$ is the sum of pair-wise interactions of each node. For $\delta = 0$ the selection probability is same for all nodes, while increasing $\delta$ makes it more likely that a newcomer chooses a node with a higher payoff. After the newcomer adopts the strategy of the chosen role-model, it also connects with the role-model with a probability $p$ and with each of the role-model’s neighbours with a probability $q$. The parameters $p$ and $q$ are called {\em embedding parameters} as they determine the ability of the newcomer to imitate the role-model’s social network. 

More importantly, the embedding parameters $p$ and $q$ also determine the stability of cooperation \citep{cavaliere2012prosperity}.  In particular, high levels of $q$ rise the risk of collapse of cooperation (level of $p$ is less relevant, Figure $S2$ in the Supplementary Material). To study how gradually increasing $q$ leads to collapse and slowly erodes the stability of cooperation, we perform a series of perturbation experiments. A perturbation experiment is defined as the introduction of an invader (e.g. cheater) in a network where all agents have the opposite strategy (e.g. cooperators). 

\begin{figure}[!htb]
	\setlength{\abovecaptionskip}{0pt}
	\setlength{\belowcaptionskip}{0pt}
	\centering 
	\includegraphics[width=1\linewidth]{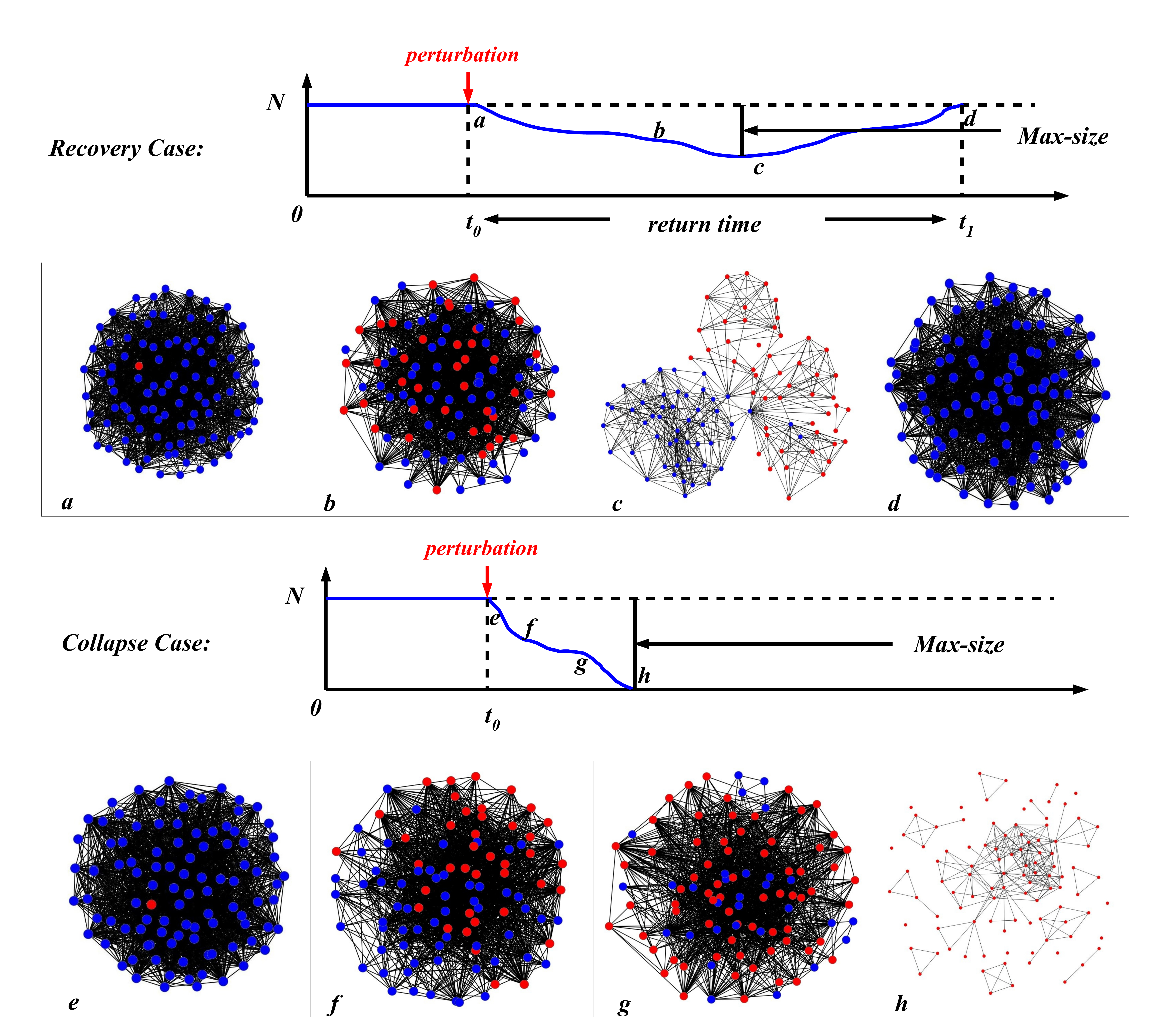}
	\caption{{\bf Recovery and Collapse of Cooperation in Perturbation Experiments.} Recovery and collapse of cooperation following the addition of a single invader in a network with agents of the opposite strategy. We show the two possible outcomes of a perturbation obtained by adding a cheater in a network of all cooperators. The first row shows the typical stages (and network topologies) of an unsuccessful perturbation where cooperators resist, while the bottom row shows the typical stages (and network topologies) of a successful perturbation where cheaters invade (in both cases $q=0.8$). Note the distinct network structure in community of all cooperators (highly connected) and in a community of all cheaters (highly fragmented)}
	\label{fig:perturbation}
\end{figure}

We start perturbations in a network of all cooperators after we have iterated the model to establish a stable network topology. We then introduce a cheater and we monitor the system until either the cheater fails to invade, {\em recovery}, or the cheater invades and cooperation collapses, {\em collapse}  (Figure \ref{fig:perturbation}). Using this approach, we evaluate the fraction of perturbations that lead to the collapse for increasing values of the embedding parameter $q$ at different selection strengths $\delta$.

\begin{figure}[!htb]
	\setlength{\abovecaptionskip}{0pt}
	\setlength{\belowcaptionskip}{0pt}
	\centering 
	\subfigure{
		\includegraphics[width=1\linewidth]{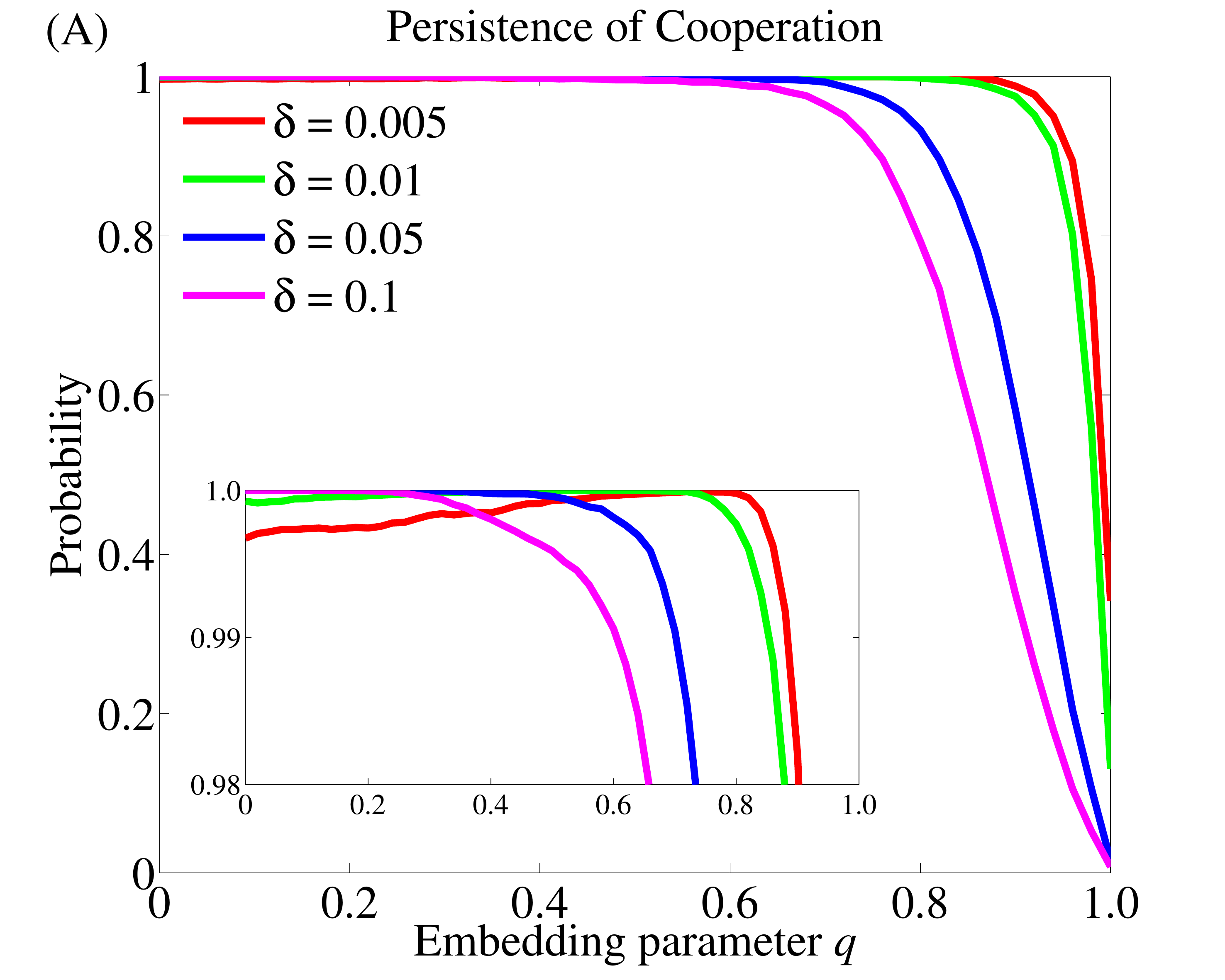}
	}
	\subfigure{
		\includegraphics[width=1\linewidth]{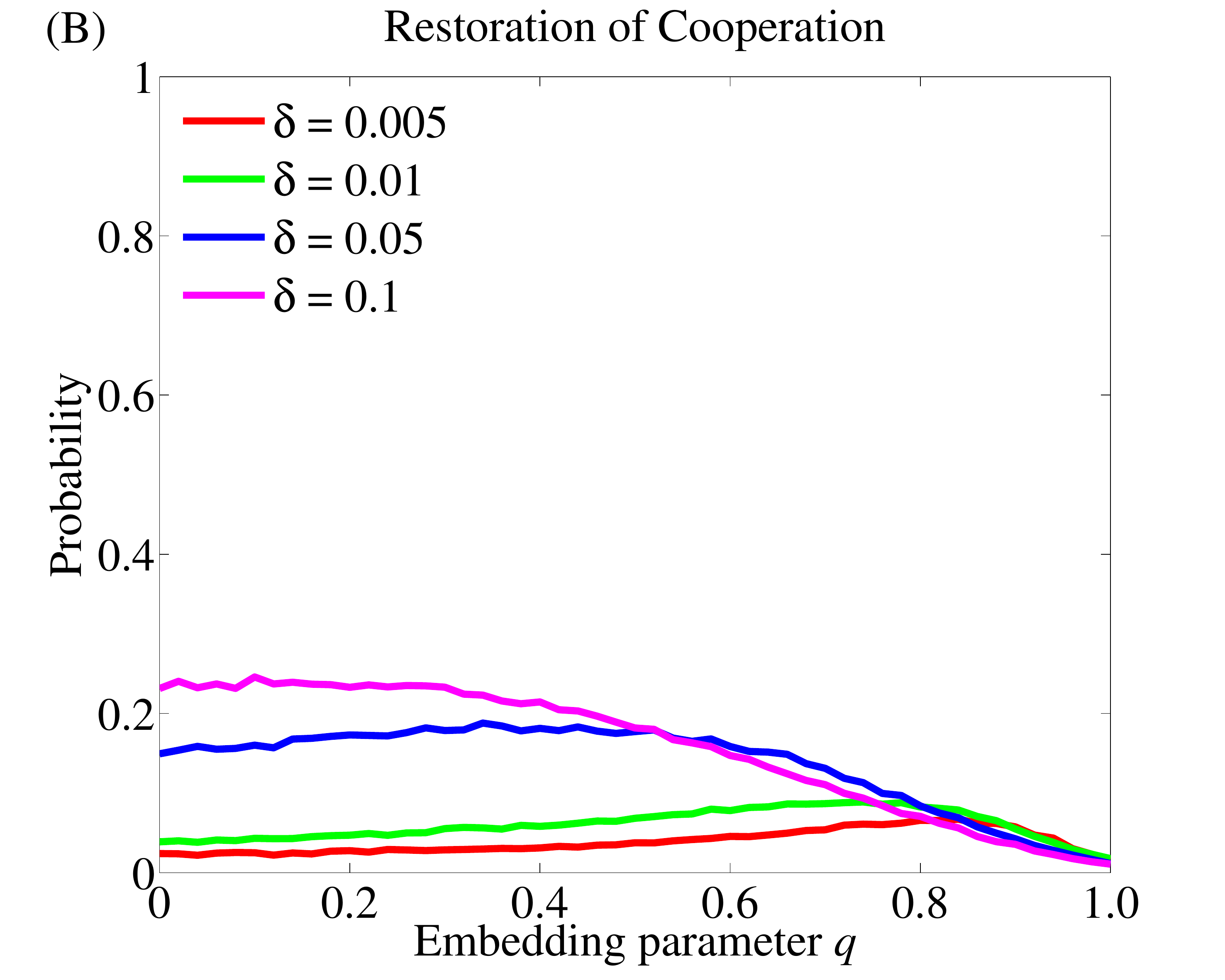}
	}
	\caption{{\bf Persistence and Restoration of Cooperation}. (A) Persistence of cooperation as a function of the embedding parameter $q$ for various selection strengths $\delta$. Note the non-monotonicity of the curves (visible clearly in the inset panel): for a low selection $\delta$, the persistence of cooperation reaches a maximum before collapsing. As selection becomes stronger, the persistence of cooperation decreases monotonically. (B) Restoration of cooperation as a function of the embedding parameter $q$ for various selection strengths $\delta$. Persistence probabilities are computed as $1-\psi$ and restoration probabilities are computed as $\psi$, where $\psi$ is the fraction of perturbations that lead to the successful invasion of the mutant out of $20.000$ perturbation experiments. Each perturbation is done by updating the network of (A) all cooperators or (B) all cheaters for a long time following the addition of a mutant at $t_0$. (All results are shown at $p=0.6$; see Supplementary Material Figure $S2$ for different $p$s).}
	\label{fig:fraction} 
\end{figure}

 This fraction represents the probability $\psi$ of the introduced cheater to spread and trigger the collapse of cooperation. We call the complement of this probability {\em cooperation persistence} (1-$\psi$). We find that cooperation stays uninvadable for a wide range of the control parameter $q$. Only when $q$ crosses a certain threshold, persistence decreases rapidly (Figure \ref{fig:fraction}A). Cooperation is doomed to fail and its loss occurs in an abrupt non-linear way. Nonetheless, the exact value of the threshold depends on the selection strength $\delta$. Higher $\delta$s cause an earlier but smoother collapse of cooperation, and turn the probability that a single cheater invades successfully almost equal to $1$ (Figure \ref{fig:fraction}A).

Once cooperation collapses, its restoration by re-introducing a cooperator in a network of cheaters is difficult. In this reverse scenario, we find that although the probability of a single cooperator to invade a network of cheaters increases for decreasing the embedding parameter $q$, it never reaches $1$ (Figure  \ref{fig:fraction}B). This means that once cooperation is lost it is difficult to be restored. This indicates that the cheaters dominance equilibrium is much more resilient than the cooperators dominance equilibrium. Restoring cooperation is only possible by inducing a stronger perturbation: a simultaneous invasion of more than a single cooperator (Figure $S1$ in the Supplementary Material). 

\subsection{Structural and Non-structural Indicators for Detecting the Collapse of Cooperation} \label{perturbations}

The abrupt loss of cooperation raises the question whether it is possible to detect it in advance. Measuring the strength of parameter $q$ is not only difficult in practice, but also largely uninformative as for a range of low $q$, as there is little sign of change in the persistence of cooperation (Figure \ref{fig:fraction}). Thus, alternative diagnostic tools would be desirable in order to estimate the rising risk of cooperation collapse. Motivated by the possibility to detect upcoming critical transition using generic indicators of resilience \citep{scheffer2009early}, we explore whether we can use similar indicators as precursors of the eroding stability of cooperation.

In particular, we evaluate two broad classes of indicators: those based on the dynamics of the community composition (that we call \emph{non-structural}) and those based on the community structure (that we call \emph{structural}). The difference between the two classes is that non-structural indicators reflect changes in the numbers of cheaters and cooperators (demographic changes), whereas structural indicators reflect changes in the interactions between cheaters and cooperators (topological changes). Among a variety of metrics, we estimate two non-structural indicators: a) return rate, the inverse of the time it takes for the system to recover back to its original state of full cooperation after the addition of a single cheater (recall that time is measured in update steps in our model), and b) maximum size, the maximal amount of cheaters possible to invade after the introduction of a single cheater. We also estimate two structural indicators: a) average degree (or connectance), the average number of links of each agent in the network, and b) the relative amount of cooperative interactions $\sigma^*$, that is the fraction of beneficial interactions between cooperators over the exploitative interactions between cooperators and cheaters (see Methods for details). Other estimated metrics are presented in the Supplementary Material (Section $7$).

We estimate these four indicators after each perturbation experiment at increasing values of the embedding parameter $q$ and for different selection strengths $\delta$. In Figure \ref{fig:signal}, we plot the median values from $20.000$ perturbations performed at each $q$ (see Supplementary Material Figure $S6$ for the distributions of all indicators). We find that both non-structural and structural indicators change in distinct ways before cooperation collapses. In particular, we compare trends among indicators up to the value of $q$ at which the probability of cooperation persistence drops below $0.5$. We use this threshold value as the most conservative choice for the onset of cooperation collapse. We observe the strongest changes for all indicators close to the transition except for average degree when selection strength $\delta$ is high (Figure \ref{fig:signal}D). In general, strong selection ($\delta=0.1$) leads to more pronounced and earlier changes in the trends when compared to weak selection (see Figure $S3$ in the Supplementary Material for different selection strengths). Perhaps most interestingly, return rate decreases before that collapse threshold (Figure \ref{fig:signal}A,C). As decreasing return rates are signature of critical slowing down prior to local bifurcation points \citep{van2007slow}, this finding indirectly supports that the loss of cooperation in evolving networks might be classified as a critical transition.

\begin{figure}[!htb]
	\setlength{\abovecaptionskip}{0pt}
	\setlength{\belowcaptionskip}{0pt}
	\centering 
     \includegraphics[scale=0.14]{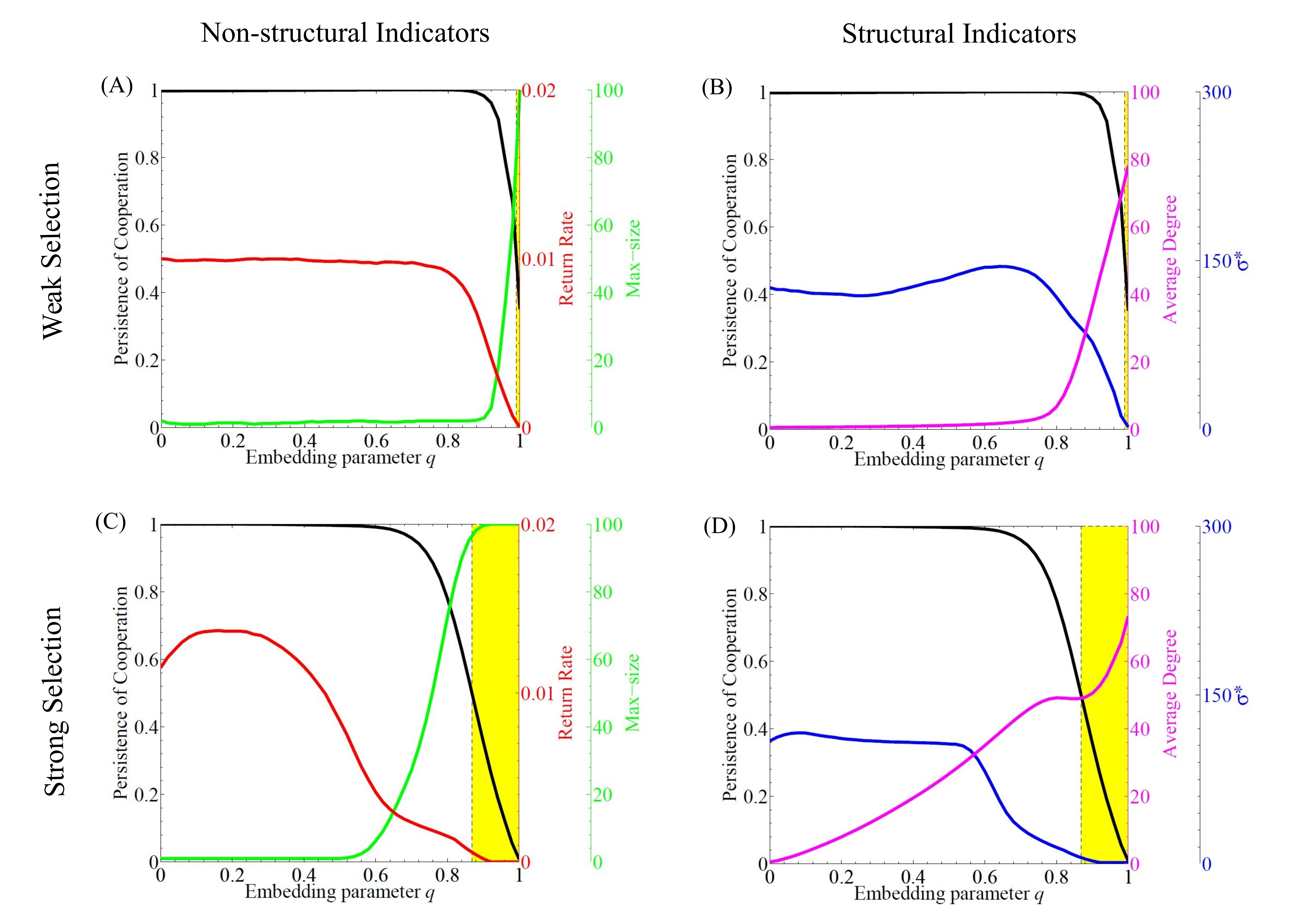}
	\caption{{\bf Detecting the Loss of Cooperation.} 
		Structural and non-structural indicators for detecting the loss of cooperation after the invasion of a single cheater at increasing levels of embedding parameter $q$. Non-structural indicators are 1) return rate (inverse of the number of update steps the system takes to go back to its original state following a perturbation), and 2) max-size (the maximal number of cheaters recorded during a perturbation). Structural indicators are 1) structural coefficient $\sigma^*$ that evaluates the ratio between the cooperative interactions and the exploitative interactions, and 2) average degree (the average number of links per node). Upper row: weak selection, $\delta=0.005$; lower row: strong selection, $\delta=0.1$. The black curves denote cooperation persistence. The yellow shaded area identifies the values of $q$ where cooperation persistence falls below $0.5$ (our defined threshold for cooperation collapse). Each point of the indicators is the median estimated out of the considering $20.000$ perturbation experiments.}
	\label{fig:signal}
\end{figure}

\subsection{Assessing the Consistency and Early Detection of the Indicators}
\begin{figure}[!htb]
	\setlength{\abovecaptionskip}{0pt}
	\setlength{\belowcaptionskip}{0pt}
	\centering 
	\subfigure{
		\includegraphics[width=1\linewidth]{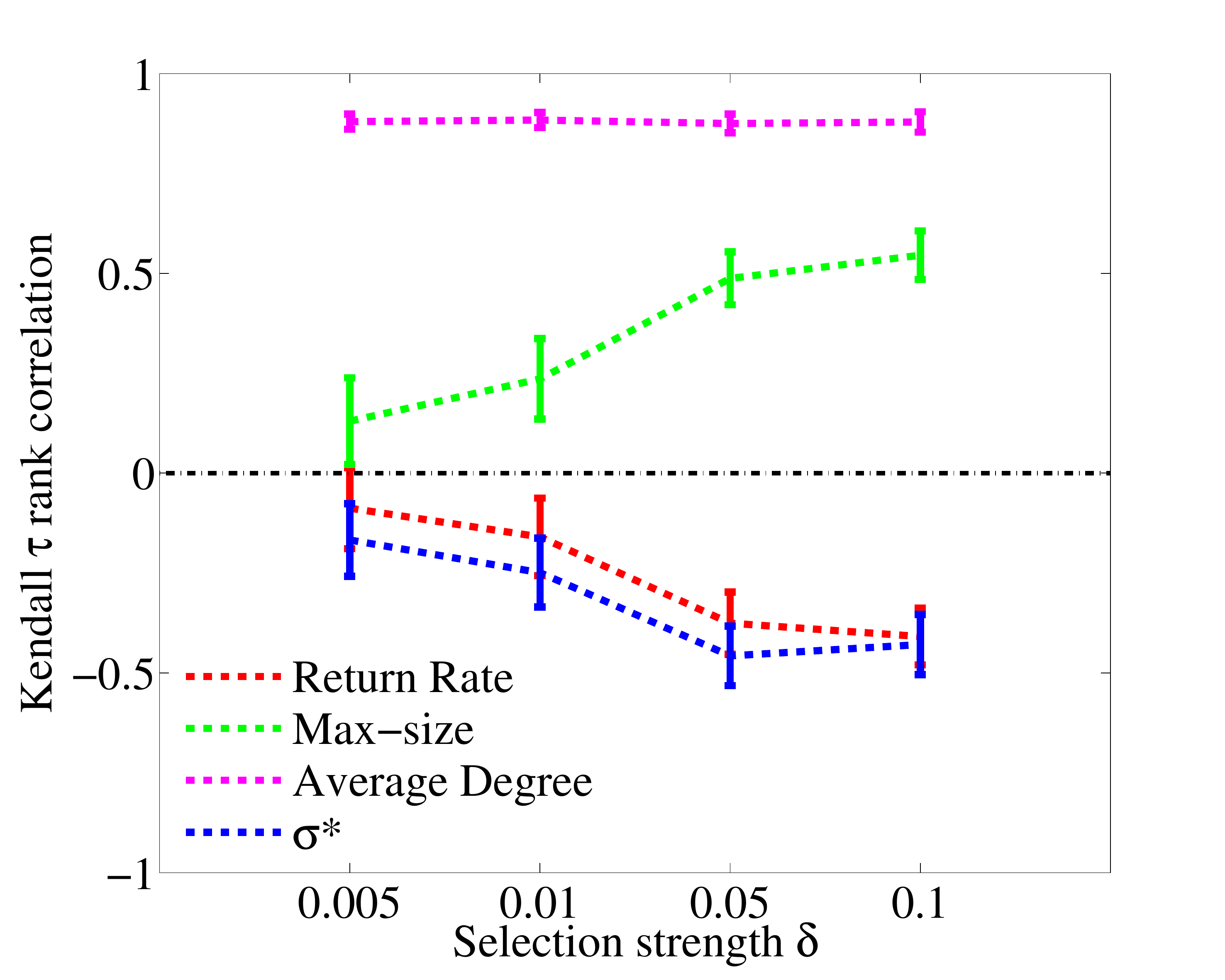}
	}
	\caption{{\bf Consistency of the indicators at different selection strengths.} Mean values and standard deviation values of the distributions for Kendall $\tau$ rank coefficients for increasing selection strengths $\delta$. Non-structural indicators are 1) return rate and 2) max-size; and the structural indicators are 1) structural coefficient $\sigma^*$ and 2) average degree. The average degree appears to be the most consistent indicator as its distribution is almost always positive across all selection strengths. In general, the consistence strength is enhanced as the selection increases.}
	\label{fig:kendall}
\end{figure}

Trends in the indicators are useful only if they are consistently correlated to changes of the embedding parameter $q$ independently from the strength of selection $\delta$. We test for such consistency by computing trends in the indicators up to the threshold where the cooperation persistence drops below $0.5$ (Figure \ref{fig:signal}). We first generate sequences of indicator values by randomly selecting a value from any of the perturbations at each $q$. Second, we computed the correlation (quantified by Kendall $\tau$ rank coefficient) between the constructed indicator sequences and the values of $q$ (see Methods). By comparing mean values of the estimated distributions of Kendall $\tau$s, we found that the trends in indicators are generally consistent across different selection strengths (Figure \ref{fig:kendall}): the mean values keep the same signs. However, when comparing the means and standard deviations of the estimated distributions of Kendall $\tau$s, we found that return rate, max-size  and $\sigma^*$ are stronger when selection strength $\delta$ increases. Average degree is the strongest indicator regardless of whether the selection is weak or strong (Figure \ref{fig:kendall}).

\begin{figure}[!htb]
	\setlength{\abovecaptionskip}{0pt}
	\setlength{\belowcaptionskip}{0pt}
	\centering 
	\subfigure{
		\includegraphics[width=1\linewidth]{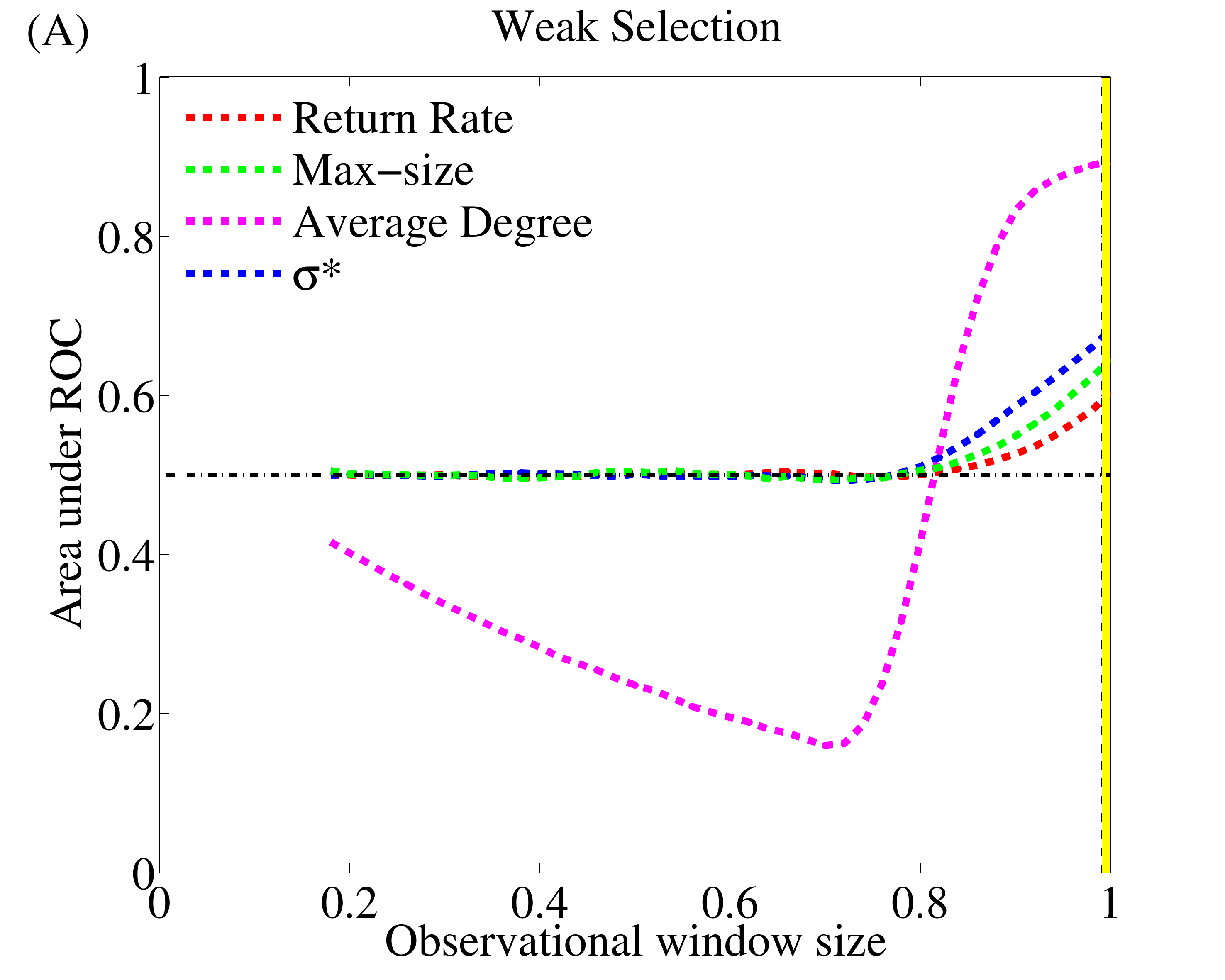}
	}
	\subfigure{
		\includegraphics[width=1\linewidth]{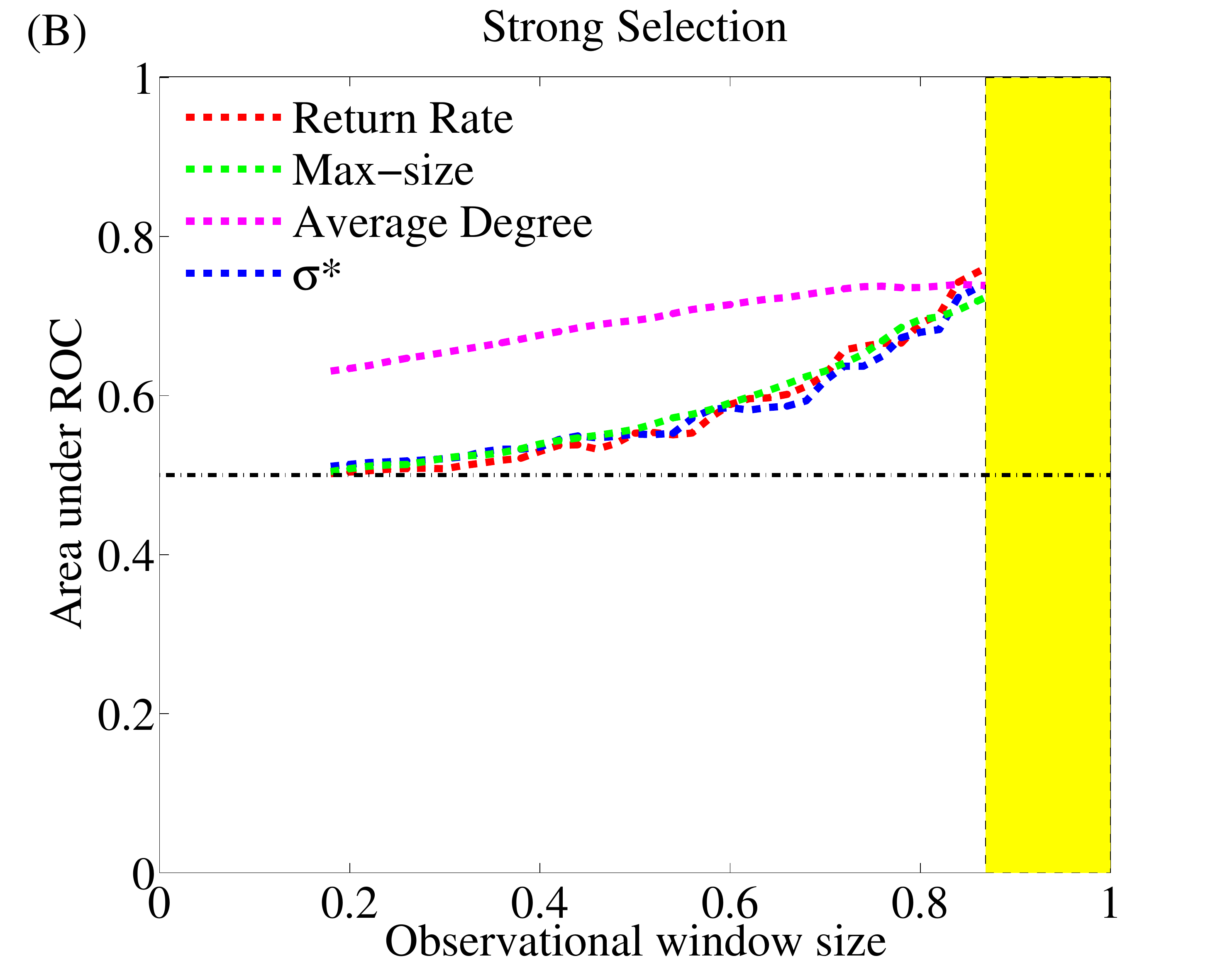}
	}
	\caption{{\bf Accuracy in the early detection of the indicators.} We plot the area under the receiver operator characteristic (ROC) curves (AUC) for structural and non-structural indicators at weak and strong selections ($\delta = 0.005$ and $0.1$) with the change of observational window. Non-structural indicators are 1) return rate and 2) max-size, and the structural indicators are 1) structural coefficient $\sigma^*$ and 2) average degree. An observational window defines the distance from the transition: a large window corresponds to comparing changes in the indicators including values close to the collapse of cooperation. An AUC of $1$ represents a perfect test; an area below $0.5$ represents a worthless test. When the selection strength is weak, the average degree displays misleading prediction information as the window size increases. In this scenario, the structural coefficient  $\sigma^*$ is generally the most accurate indicator. At strong selection, all indicators are more accurate than those at weak selection.}
	\label{fig:AUC}
\end{figure}

We also analyze how ``early" an indicator accurately signals the increasing risk of collapse. Or, in other words, how far from the collapse changes in the indicators significantly signal the rising risk of cooperation loss. We do this by using receiver operator characteristic (ROC) curves \citep{boettiger2012quantifying} that estimate true positive and false positive rates for all possible cut-off levels of the indicators (see Methods). The larger the area under the ROC curve (AUC), the more accurately an indicator identifies the risk of cheater's invasions. Areas below $0.5$ mean that the indicator trend carries no accurate information about the risk of invasion. We compare the estimated area of the ROC curves for each indicator across a range of observation window sizes (Figure \ref{fig:AUC}). The size of the observation window is inversely related to the distance from the collapse threshold. Under weak selection, all indicators are poorly signaling the increasing risk of collapse. Only at big window sizes that include estimates close to the collapse, the accuracy of the indicators increases (above $0.8$ window size).  Notably, average degree is inaccurately detecting proximity to collapse for most window sizes (Figure \ref{fig:AUC}A). When selection is strong (Figure \ref{fig:AUC}B), all indicators more accurately signal the rising risk of cooperation loss compared to the case at weak selection (Figure \ref{fig:AUC}A).

%
%
\section{Methods} \label{methods} 
\subsection{Model} \label{networkupdate}
We consider a network model with a fixed number of agents (nodes) but with a non-fixed number of links: to whom and to how many neighbours an agent is connected varies during the evolution of the system. Each agent in the network adopts one of the two strategies of the {\em Prisoner's Dilemma}. A {\em cooperator} pays a cost $c$ to provide a benefit $b$ to all of its neighbours; but the {\em cheaters} pay no cost and distribute no benefit. For instance, if a cooperator has $m$ cooperative neighbours and $n$ cheating neighbours, its payoff is $m(b-c)-nc$. A cheater in the same neighbourhood has payoff $mb$. We use $b/c=3$ in our numerical experiments. The dynamics of the system are defined by a discrete sequence of {\em update steps} (Figure \ref{fig:evolution}). At each update step a new node (a {\em newcomer}) is added and a randomly chosen existing node is removed from the system so that the number of nodes is constant (in the main text we assume $N=100$). As the newcomer has no specific strategy, a node $i$ is selected as a role-model with a probability proportional to its \emph{effective payoff} $EP_i$ = (1+$\delta)^{\emph{Payoff}_i}$, where $\delta \geq 0$ specifies the intensity of selection and $\emph{Payoff}_i$ is the sum of pair-wise interactions of each node. For $\delta = 0$ the selection probability is same for all nodes, while increasing $\delta$ makes it more likely that a newcomer chooses a node with a higher payoff. A newcomer has the same strategy of the role-model and connects with it with a probability $p$ as well as with each of its neighbours with a probability $q$ (that means $q^k$ is the probability that a newcomer connects to all $k$ neighbours of the role-model). The parameters $p$ and $q$ are called {\em embedding parameters} as they explicitly determine the ability of the newcomer to imitate the role-model’s social network. In particular, the embedding parameter $q$ determines the ability of newcomers to link to a high number of agents in the network.  In this paper we focus on a version of the model described in the work \citep{cavaliere2012prosperity}, where the newcomers always copy (no mutation is allowed) the strategy of the selected role-model. The degree of competition, {\em selection strength}, is controlled by the parameter $\delta$, which determines the effective payoff. The standard model  \citep{cavaliere2012prosperity} where newcomers can select a different strategy than the one used by the role-model has been analyzed in the Supplement (Section $9$).

\subsection{Evaluation of the Indicators}  \label{indicators}
We estimate two classes of indicators: structural and non-structural. Non-structural indicators consider only the population composition, while structural indicators consider the structure of the network. We compute structural and non-structural indicators for each perturbation experiment. A perturbation experiment consists of the introduction of a mutant in a network where all agents have the opposite strategy. We only compute the indicators in the scenario of the invasion of a single cheater in a network of all cooperators. This is done with the following procedure. A network of all cooperators is updated for $1000$ steps to remove transients. We then introduce a cheater newcomer. The system is updated until one of the two outcomes is reached: either recovery, the cheater fails to invade, or collapse, the cheater invades and cooperation collapses. We perform $20.000$ of these perturbation experiments for each value of the embedding parameter $q$ increasing from $0$ to $1$ at an increment of $0.02$. We use different selection strengths ($\delta =0.001, 0.005, 0.01, 0.05$), but in the main text we report the indicator results for weak ($\delta = 0.005$) and strong selection strength ($\delta= 0.1$). We identify by $t_0$ the beginning of a perturbation (the addition of the cheater) and with $t_{end}$ the end of a perturbation (either recovery or collapse). For each perturbation experiment, we compute the indicators as follows regardless of whether a perturbation will lead to recovery or collapse.

\begin{itemize}
	\item Non-structural indicators
	
	\subitem - Return rate is $\frac{1}{\emph{Return time}}$, where the \emph{Return time} is the number of update steps the system takes to go back to its original state following a perturbation.  Hence, if $t_0$ is the step at which a perturbation starts and  $t_{end}$ is the step at which the population comes back to its original state, then the return time is defined as $t_{end}- t_0$.  If the perturbation is successful, then the return rate is defined to be $0$. 
	
	\subitem - Max-size (of cheaters) is the maximal number of cheaters recorded during a perturbation to a network of cooperators. It is the size $N$ of the population if the perturbation leads to collapse.
	
	\item Structural indicators.
	\subitem - Structural coefficient $\sigma^*$:
	\begin{equation}
	\sigma* = \frac{\sum_{t=t_0}^{t_{end}} [CC]_t}{\sum_{t=t_0}^{t_{end}} [CD]_t}
	\end{equation}
	
	where $[CC]_t$ is the total number of $CC$ links (counted twice), $[CD]_t$ is the total number of $CD$ links in the network at step $t$. Intuitively  $\sigma^*$ evaluates the ratio between the beneficial payoff (generated by purely cooperative interactions) and the detrimental payoff (generated by cheaters connected to cooperators). This coefficient is a simplified version of the structural parameter studied in the work \citep{tarnita2009strategy, nowak2010evolutionary}. 
	
	\subitem - Average degree is the average number of links per node recorded during a perturbation to a network of cooperators.
\end{itemize}

\subsection{Consistency of Indicators - Kendall's $\tau$ coefficient} \label{consistency}

Given two pairs of data, e.g., $(x_i, y_i)$ and $(x_j,y_j)$, we say they are \emph{concordant} if $x_i>x_j$ and $y_i>y_j$ or if $x_i<x_j$ and $y_i<y_j$; otherwise if $x_i>x_j$ and $y_i<y_j$ or $x_i<x_j$ and $y_i>y_j$ then we say they are \emph{discordant}. Suppose we have two sequences (corresponding to two variables), each with $n$ data points, the Kendall's $\tau$ coefficient between the two variables is computed using the number of concordant (cp) and discordant pairs (dp):
\begin{equation}
\tau = \frac{cp - dp}{n(n-1)/2}
\end{equation}

If the agreement between the two variable is perfect the coefficient has a value of $1$; if the disagreement is perfect then it has a value of $-1$; while if the two variables are independent then it is $0$. The coefficient can be interpreted as the probability of observing concordant pairs minus the probability of observing discordant pairs \citep{conover1980practical}.

Let's denote $q(x)$ as the value of $q$ for which the persistence of cooperation is $x$. For each indicator, we compute the Kendall's $\tau$ coefficient by constructing indicator sequence whose length is the number of $q$s considered between $q=0$ and $q(0.5)$ and each corresponding value is randomly chosen from $20.000$ perturbations. In this way, a large number of sequences can be constructed until all data are exhausted following a bootstrap approach without replacement. The distribution is fitted as a Gaussian one (Figure $S4$ in the Supplementary Material) through those obtained the Kendall $\tau$ coefficients between each of these constructed indicator sequences and the corresponding embedding parameter $q$s.

\subsection{Accuracy of Indicators - ROC curves} \label{roc}
The observational window is from $q=0$ to $q'$, and $q'\leq q(0.5)$. For each $q'$ in Figure \ref{fig:AUC}, we plot the area under the ROC curve (AUC) that is computed using all (and only) the data between $q=0$ and $q'$. Specifically, for each indicator we compute two distinct ROC curves in the following manner ---that correspond to two possible discrimination conditions.

We use a cut-off $c$ to define true/false positive/negatives as follows. Given two arbitrary points $q_1,q_2$ between $q=0$ and $q'$, with $q_1<q_2$, we evaluate the values of the indicator at $q_1$ and at $q_2$ and denote them by $s_1$ and $s_2$, respectively. We denote by $f(q_1)$ and $f(q_2)$ the values of the cooperation persistence at $q_1$ and $q_2$, respectively. Consequently, we can define the true/false positive/negative by considering two possible (symmetric) discriminatory conditions.

Condition (I):
\begin{itemize}
	\item If $(1+c)s_1<s_2$ and $f(q_1)\geq f(q_2)$, then it is a true positive (TP).
	\item If $(1+c)s_1<s_2$ and $f(q_1)< f(q_2)$, then it is a false positive (FP).
	\item If $(1+c)s_1\geq s_2$ and $f(q_1)< f(q_2)$, then it is a true negative (TN).
	\item If $(1+c)s_1\geq s_2$ and $f(q_1)\geq f(q_2)$, then it is a false negative (FN).
\end{itemize}

Condition (II):
\begin{itemize}
	\item If $(1+c)s_1\geq s_2$ and $f(q_1)\geq f(q_2)$, then it is a true positive (TP).
	\item If $(1+c)s_1 \geq s_2$ and $f(q_1)< f(q_2)$, then it is a false positive (FP).
	\item If $(1+c)s_1< s_2$ and $f(q_1) < f(q_2)$, then it is a true negative (TN).
	\item If $(1+c)s_1< s_2$ and $f(q_1)\geq f(q_2)$, then it is a false negative (FN).
\end{itemize}

For a given cut-off $c$, we can produce a large number of random indicator pairs $(s_1, s_2)$ and corresponding cooperation pairs $(c_1, c_2)$, which are picked between $q=0$ and $q'$. And then we can evaluate, for each indicator, the total number TP, FP, TN and FN, using either condition (I) or condition (II). Once these values have been obtained, we compute the sensitivity and specificity for each indicator as:
\begin{equation}
\mbox{Sensitivity} = \mbox{true positive rate} = \frac{\# TP}{\# TP + \# FN}
\end{equation}
\begin{equation}
\mbox{1-Specificity} = \mbox{false positive rate} = \frac{\# FP}{\# TN + \# FP}
\end{equation}
The calculated pair (sensitivity, 1-specificity) constitutes a single point of the ROC curve. Repeating the described process for all possible cut-offs $c$ (from very small to very large), we can obtain a set of pairs (sensitivity, 1-specificity) that constitute a full ROC curve.

Each indicator has two possible symmetric ROC curves, depending whether the condition used to determine FP, FN, TP, TN is either (I), or (II). For convention, we choose the one that gives the largest AUC when computed within the largest observational window, and use that condition to compute the ROC for all other observational windows. For completeness, we report the ROC curves in the Supplementary Material (Figure $S5$). 

\section{Discussion} \label{discussion}

Recognizing the conditions that favor the spreading of cheating and the collapse of cooperation in a community has been a major objective in the study of complex adaptive systems\citep{levin1999fragile, levin2010crossing, bowles2011cooperative}. Here we approach this issue from a different  perspective and look at the dynamics of cheaters, cooperators, and their interactions to infer the risk of cooperation collapse. Specifically, we argue that the collapse of cooperation resembles a critical transition in our structurally evolving communities - an abrupt change in equilibrium state at the crossing of a threshold. Although, it is difficult to analytically show the nature of this transition, we find that return rate (i.e., the inverse of the time necessary for a cheater to get expelled from the community) decreases prior to the collapse in our numerical experiments (Figure \ref{fig:signal}). Slow return rates resemble the critical slowing down that is hallmark of proximity to bifurcations points \citep{scheffer2009early, wissel1984universal}, but it clearly appears in our evolving complex communities. 

Apart from decreasing return rates, we found that the abrupt loss of cooperation can be detected in advance by a handful of structural as well as non-structural indicators that are not linked to critical slowing down (Figure \ref{fig:signal}). We show that the maximum size of invaders is increasing relatively to the risk of collapse. Intuitively, this indicator reflects the disturbance (in terms of invader numbers) that the community can tolerate without collapsing. On the other hand, we use, for the first time, structural indicators for detecting instabilities in network dynamics. The few studies on critical transitions in networks assume fixed topologies and use non-structural indicators to detect species extinctions \citep{dakos2015resilience, suweis2014early}. Network metrics, like connectance, assortativity, or clustering, have been used as structural indicators applied on spatial dynamical models where network structure was derived from cross-correlation matrices \citep{tirabassi2014interaction, viebahn2014critical}. However, these matrices were static. In our communities, network structure is not static but it co-evolves with the spreading of strategies depending on the expected differences in payoff between cheaters and cooperators. Clearly, the dynamic nature of the structure of our communities permits to test indirect alternative metrics for estimating the risk of abrupt transitions. Among the structural indicators we tested (Figure \ref{fig:signal} and Figure $S7$ in the Supplementary Material), we show that the decreasing ratio of beneficial interactions (between cooperators) to detrimental ones (between cheaters and cooperators) can be successfully used to detect the increasing risk of cheater's invasions (Figure \ref{fig:signal}). Moreover such indicator is the most accurate when the strength of selection is weak (Figure \ref{fig:AUC})

Tests about the consistency and accuracy of the estimated trends (Figure \ref{fig:kendall} and \ref{fig:AUC}) strongly imply that there is high uncertainty in detecting robustly the loss of cooperation in an evolving community. Such uncertainty in the detection of critical transitions of resilience has been reported to be caused by a variety of factors (e.g. heterogeneity, non-linear dynamics, measurement error, strong environmental stochasticity \citep{dakos2015resilience,boettiger2013early}). In our communities, we found that detection is strongly dependent at least on the underlying level of selection strength. This does not come as a surprise. For weaker selection strength, cooperators are less disfavored allowing them to persist even at high levels of the embedding parameter $q$. This implies that all indicators, but average degree, can ``pick up" the increase of fragility only at very high values of $q$ close to the collapse (Figure \ref{fig:kendall}).

Clearly, our approach is limited as we considered the dynamics of the communities after a single invasion. A more realistic scenario would allow for constant invasions of cheaters, or mutations of cheaters to cooperators as well. In that case, cooperation would collapse and recover in an infinite sequence of such events \citep{cavaliere2012prosperity}. When we tested such more realistic scenario of long-term evolution with endogenous mutations, we found that most of the indicators could still detect the increasing risk of cooperation loss (Section $9$ in the Supplementary Material). Similarly, although we only tested a handful of indicators, it appears that alternative metrics could also be used as potential indicators to detect the cooperation loss (Figure $S7$ in the Supplementary Material). We also found that some of the considered indicators (e.g. average degree) can detect the collapse of cooperation even if in the absence of cheater invasions (Figure $S8$ in the Supplementary Material). 

Despite our work is based on a specific dynamical network model, the approach and the indicators we developed can be easily extended to other evolutionary models of dynamical structured populations \citep{lieberman2005evolutionary, tarnita2009strategy, tarnita2009evolutionary}. In these models, it is straightforward to measure the same or similar indicators and to evaluate their consistency. For example, average degree appears to be a good indicator when the selection is either weak or strong, but it is the relative number of interactions between cooperators and cheaters that accurately detects whether cheaters have better (or worse) chances to invade (Figure \ref{fig:AUC}). Similar features of the indicators may be presented in other models of evolving structured populations, where the structural coefficients have been shown to be key for studying the long-term level of cooperation \citep{tarnita2009strategy, nowak2010evolutionary}. 

The presented results suggest that it is possible to evaluate the fragility of cooperation by monitoring interactions and following the behaviour of individuals in a community. In practice, however, such information might be difficult to obtain. Identifying cheaters from cooperators, or even following their interactions in time seems a daunting task. Still, the difficulty for evaluating the proposed signals may depend on the specific application, the type, and the scale of the community in question.  For instance,  bacterial  communities are emerging as a promising experimental tool to validate hypotheses in ecology and evolution \citep{velenich2012synthetic}. Testing the proposed indicators may be  done with cooperative cells that produce public goods. In such bacterial colonies, in principle one could perform a series of perturbation experiments by introducing cells of cheaters (or invaders) and  measuring  return rate.  Recent papers have showed how structured cellular communities can be used to measure critical slowing down in deteriorating populations \citep{dai2013slower}, even in the presence of cheaters exploiting public good producing yeast \citep{chen2014dynamics}.

Perhaps the most novel conclusion of this work is that we can identify the progressive loss of cooperation by combining non-structural and structural indicators. Although the patterns we report might be idiosyncratic to our model assumptions, they still confer a promising pattern. As such our study paves the way for testing and developing similar indicators in a variety of evolving dynamical networks, ranging from biological systems to ecological communities and even socio-economic networks.

\section{Acknowledgements}

We acknowledge helpful discussions with C.E. Tarnita on the structural indicators. V. Dakos was supported by a MarieCurie IntraEuropean postdoctoral fellowship and M. Cavaliere was supported by Engineering and Physical Sciences Research Council grant number EP/J02175X/1. 

\section{Author contributions statement}
M. Cavaliere. and V.Dakos conceived the study and wrote the paper. G.Yang conducted the computational experiments. All authors analysed the results and reviewed the paper. 

\bibliographystyle{aaa}
\bibliography{ref}

\end{document}